\documentclass[12pt]{article}
\usepackage{amsfonts}
\usepackage{amssymb,amsmath}
\newcommand{\ol}{\setlength{\itemsep}{0pt.}\begin{enumerate}}
\newcommand{\eol}{\end{enumerate}\setlength{\itemsep}{-\parsep}}

\newtheorem{THEOREM}{Theorem}[section]  
\newenvironment{theorem}{\begin{THEOREM} \hspace{-.85em} {\bf :} 
}%
                        {\end{THEOREM}}
\newtheorem{LEMMA}[THEOREM]{Lemma}
\newenvironment{lemma}{\begin{LEMMA} \hspace{-.85em} {\bf :} }%
                      {\end{LEMMA}}
\newtheorem{COROLLARY}[THEOREM]{Corollary}
\newenvironment{corollary}{\begin{COROLLARY} \hspace{-.85em} {\bf 
:} }%
                          {\end{COROLLARY}}
\newtheorem{PROPOSITION}[THEOREM]{Proposition}
\newenvironment{proposition}{\begin{PROPOSITION} \hspace{-.85em} 
{\bf :} }%
                            {\end{PROPOSITION}}
\newtheorem{DEFINITION}[THEOREM]{Definition}
\newenvironment{definition}{\begin{DEFINITION} \hspace{-.85em} {\bf 
:} }
                            {\end{DEFINITION}}
\newtheorem{EXAMPLE}[THEOREM]{Example}
\newenvironment{example}{\begin{EXAMPLE} \hspace{-.85em} {\bf :} 
\rm}%
                            {\end{EXAMPLE}}
\newtheorem{CONJECTURE}[THEOREM]{Conjecture}
\newenvironment{conjecture}{\begin{CONJECTURE} \hspace{-.85em} 
{\bf :} \rm}%
                            {\end{CONJECTURE}}
\newtheorem{PROBLEM}[THEOREM]{Problem}
\newenvironment{problem}{\begin{PROBLEM} \hspace{-.85em} {\bf :} 
\rm}%
                            {\end{PROBLEM}}
\newtheorem{REMARK}[THEOREM]{Remark}
\newenvironment{remark}{\begin{REMARK} \hspace{-.85em} {\bf :} 
\rm}%
                            {\end{REMARK}}
\newtheorem{CONCLUSION}[THEOREM]{Conclusion}
\newenvironment{conclusion}{\begin{CONCLUSION} \hspace{-.85em} {\bf :} 
\rm}%
                            {\end{CONCLUSION}}
\newcommand{\thm}{\begin{theorem}}
\newcommand{\lem}{\begin{lemma}}
\newcommand{\pro}{\begin{proposition}}
\newcommand{\dfn}{\begin{definition}}
\newcommand{\rem}{\begin{remark}}
\newcommand{\con}{\begin{conclusion}}
\newcommand{\xam}{\begin{example}}
\newcommand{\cnj}{\begin{conjecture}}
\newcommand{\prb}{\begin{problem}}
\newcommand{\cor}{\begin{corollary}}
\newcommand{\prf}{\noindent{\bf Proof:} }
\newcommand{\ethm}{\end{theorem}}
\newcommand{\elem}{\end{lemma}}
\newcommand{\epro}{\end{proposition}}
\newcommand{\edfn}{\bbox\end{definition}}
\newcommand{\erem}{\bbox\end{remark}}
\newcommand{\econ}{\bbox\end{conclusion}}
\newcommand{\exam}{\bbox\end{example}}
\newcommand{\ecnj}{\bbox\end{conjecture}}
\newcommand{\eprb}{\bbox\end{problem}}
\newcommand{\ecor}{\end{corollary}}
\newcommand{\eprf}{\bbox}
\newcommand{\beqn}{\begin{equation}}
\newcommand{\eeqn}{\end{equation}}

\newcommand{\bbox}{\vrule height7pt width4pt depth1pt}

\def\hook{\hbox{\vrule height0pt width4pt depth0.3pt
\vrule height7pt width0.3pt depth0.3pt \vrule height0pt width2pt
depth0pt} }
\def\sqr#1#2{{\vcenter{\hrule height.#2pt\hbox{\vrule width.#2pt
height#1pt \kern#1pt \vrule width.#2pt}\hrule height.#2pt}}}

\newcommand{\sect}[1]{\setcounter{equation}{0}\bigskip\medskip
\section{#1}\smallskip}

\def\br{\begin{eqnarray}}
\def\er{\end{eqnarray}}
\def\brn{\begin{eqnarray*}}
\def\ern{\end{eqnarray*}}
\def\er{\end{eqnarray}}
\def\eqq{&\!\!\!\!\!=\!\!\!\!\!&}
\def\eee{&\!\!\!\!\!\!\!\!\!\!&}
\def\vt{\vartheta}

\def\L{{\cal{L}}}

\def\a{\alpha}
\def\b{\beta}

\def\d{\delta}

\def\la{\lambda}
\def\L{\mathcal{L}}

\newcommand{\bi}[1]{\bibitem{#1}}
\title{{\bf On variations in teleparallelism theories}}
\author{
\thanks {\quad   email itin@sunset.ma.huji.ac.il}
Yakov Itin \\
\small {Institute of Mathematics}\\
\small {Hebrew University of Jerusalem}\\
\small {Givat Ram, Jerusalem 91904, Israel}\\
}
\begin{document} 
\maketitle
\begin{abstract}
\sf
The variation procedure on a teleparallel manifold is studied.
The main problem is the non-commutativity of the variation with the Hodge 
dual map. We establish certain useful formulas for variations and restate 
the master formula due to Hehl and his collaborates. Our approach is different 
and sometimes easier for applications. By introducing the technique of the 
variational matrix we find necessary and sufficient conditions for commutativity 
(anti-commutativity) of the variation derivative with the Hodge dual operator. 
A general formula for the variation of the quadratic-type expression is obtained. 
The described variational technique are used in two viable field theories: 
the electro-magnetic Lagrangian  on a curved manifold and the Rumpf Lagrangian 
of the translation invariant gravity. 
\end{abstract}
\sf
\sect{Introduction}
One of the biggest challenges of the theoretical physics is to unify
the standard model of particles interactions with Einstein's theory
of gravity. Both these theories are in a good
according with all known observable phenomena.
But their fundamental concept are rather differ.
Standard model is a quantum field theory on a flat Minkowskian space-time.
In contrast to that the Einstein's description of gravity is a classical
field theory essential
connected with the geometrical properties of the pseudo-Riemannian manifold.
One can expect the combination of these different theories only after
principal modification of every one of them.
Great attempts are making recently in modernization the standard model
in accordance with the idea  of superstrings. The resent description of the
situation in this area can be found in \cite{S-S}.
The second partner of the future couple should also make  some necessary
preparations.
The Einsteinian theory of general relativity (GR) is almost 
a unique  theory of gravity
that can be formulated in the framework of the pseudo-Riemannian
manifold.
Thus an essential modification can be made only on a basis of some generalized
geometrical construction. The study of various geometries relevant
to gravity began at once after Einstein had proposed his general relativity.
The most general geometrical theory of gravity is the metric-affine theory
\cite{hehl95}. In the present paper we study a rather simpler generalization of
the pseudo-Riemann geometry - the theories of teleparallelism.
The teleparallel space was introduced for the first time by Cartan \cite{Ca} 
and used by 
Einstein \cite{E1}, \cite{EM} in a certain variant of his unified 
theory of gravity and electromagnetism. 
The work of Weitzenb\"{o}ck \cite{We1} was the first 
denoted to the investigation of the geometric structure 
of teleparallel spaces. 
The theories based on this geometrical structure appear in physics 
time to time in order to give an alternative model of gravity 
or to describe the spin properties of matter. For the resent investigations 
in this area see Ref.  \cite{Kop}, \cite{M-H},  
\cite {Maluf2}, \cite {Maluf3}, \cite {T-N}, \cite {N-Y}.\\
It is convenient and useful to have a Lagrangian formulation of a field theory.
A designated functional, denoted by action, is the basis for the 
Lagrangian formalism.
The field equations of the classical field theory are the representation 
of the critical point condition for the prescribed action functional.\\
Let us take a brief look on the main properties of the action functionals 
in the classical relativistic field theory.
\begin{itemize}
\item The action functional is an integral taken on the whole 4-dimensional 
differential manifold.
\item The integrand  is a certain differential 4-form - \textsl{ Lagrangian density}\sf.
\item The Lagrangian density is a scalar invariant under the group of 
transformations of the considered system.   
\item The Lagrangian density incorporates only the squares of the 
first-order derivatives of the field variables in the same point - local densities. 
\end{itemize}
We consider the last  condition as a necessary one, 
because almost all the physically meaningful field 
equations are second-order partial differential equations of hyperbolic type.
The applying of the variational principle on the teleparallel framework is 
connected with various problems, as it emphasized in \cite{Hehl}. 
 The main source of this problems connects with the fact that the operators of 
a certain type, namely the Hodge dual operator, the interior product 
and the coderivative operator, non-commute with the variational derivative.   
In the present paper we study the the variation procedure 
(free and constrained) on the teleparallel manifold. 
The overview of the work is as following. We begin with a brief overview of 
the variational procedure in the general relativity. 
In the second section we describe 
the preferences that bring the consideration of  a teleparallel space instead 
of a pseudo-Riemannian manifold. The most advantage of such generalization is 
a possibility to describe the gravity by  the quadratic Lagrangians, similar 
to the other field theories. 
Variational procedure on the teleparallel spaces is described in the third section. 
We prove the commutative relation which coincides in the case of pseudo-orthonormal 
coframe with the master formula of 
Muench, Gronwald and Hehl \cite{Hehl}. The analogous relation described also in 
\cite{BFK}. Using the  commutative relation we derive  the formula for 
variation of the general quadratic Lagrangian. The fourth section is devoted for 
describing the various types of the covariant constrains on a teleparallel 
structure. We study the relation of such constrains with the commutativity  and 
anti-commutativity of the variational derivative with the Hodge dual map. The last 
two sections are devoted to the application of our variational formula for 
viable field theories. We consider the Maxwell Lagrangian on a teleparallel 
space and the Rumpf Lagrangian for a translation invariant gravity. 

\sect{Actions on a teleparallel  manifold.} 
The teleparallel space was considered originally as a 4D-space
endowed with  a smooth field of frame (an ordered set of 4 independent
vectors). Thus one 
can give a sense to parallelism  of two vectors taken in different
points of the space.
In order to furnish the action with a differential 4-form and to apply
the technique of the 
differential forms we consider a coframe field instead of the frame field.
Thus
\dfn
A teleparallel manifold is a pair $(M,\Gamma)$, where $M$ is a
differential 4D-manifold 
and $\Gamma=\{\vt^a(x^\mu\}, \ a=0,1,2,3$ 
is a fixed smooth cross-section of the coframe bundle $FM$.
\edfn 
The field variable is a 4-tuple of 1-forms $\vt^a$ and they constitute  
a basis of the covector space $T^*_xM$ at every point $x$ of $M$. 
We endow this vector space with the Lorentzian
metric $\eta^{ab}=\eta_{ab}=diag(-1,1,1,1)$ and
require the 1-forms $\vt^a$ to be pseudo-orthonormal
relative to this metric. Let $C^\infty$ be a set of smooth real valued
functions on $M$.
Denote by $\Omega^p$ the $C^\infty$-modulo of differential $p$-forms on $M$.\\
The following algebraic operations are defined on a $n$-dimensional 
teleparallel manifold:  
\begin{itemize}
\item[\bf{1}] The exterior (wedge)  product $\wedge:\Omega^p\times\Omega^q\to\Omega^{p+q}$ 
of two differential forms $\a\in\Omega^p$ and $\b\in\Omega^q$, which is associative, 
$C^\infty$-bilinear and in general not commutative
\begin{equation}
\a\wedge\b=(-1)^{pq}\b\wedge\a.
\label{2-1}
\end{equation}
\item[\bf{2}] The Hodge dual map $*:\Omega^p\to\Omega^{n-p}$.
This is  a $C^\infty$-linear map 
$*:\Omega^p\to\Omega^{n-p}$, which acts 
on the wedge  product monomials of the basis 1-forms as
\footnote{We use the abbreviated notations for  the wedge  product monomials:
$\vt^{ab\cdots}:=\vt^a\wedge\vt^b\wedge\cdots$}
 \begin{equation}
*(\vt^{a_1\cdots a_p})=\epsilon^{a_1\cdots a_n}\vt_{a_{p+1}\cdots{a_n}},
\label{2-2}
\end{equation}
where $\vt_a$ is the down indexed 1-forms $\vt_a=\eta_{ab}\vt^b$ 
and $\epsilon^{a_1...a_n}$ is the total antisymmetric pseudo-tensor.
\item[\bf{3}] The interior product $v\hook \a$ of an arbitrary differential form 
$\a\in \Omega^p $ with an arbitrary vector $v\in T_xM$. This is a  $C^\infty$-bilinear map 
$\hook:\Omega^p\to\Omega^{p-1}$, which satisfies the following properties:
\begin{equation}
v\hook (\a\wedge\b)=(v\hook \a)\wedge\b+(-1)^{deg\a}\a\wedge(v\hook \b),
\label{2-3}
\end{equation}
\begin{equation}
e_a\hook \vt^b=\d^b_a,
\label{2-4}
\end{equation}
where $e_a$ is a basis vector of $T_xM$ and $\vt^a$ is a basis 1-form of $T^*_xM$.
\end{itemize}
Define a  product for an arbitrary 1-form $w$ and  an arbitrary $p$-form as 
\begin{equation}
w\hook \a:=*(w\wedge*\a).
\label{2-5}
\end{equation}
It is easy to see that this operation is a $C^\infty$-bilinear map 
$\Omega^p\to\Omega^{p-1}$, which satisfies the properties (\ref{2-3}, \ \ref{2-4}). 
In particular, we have
 \begin{equation}
\vt_a\hook \vt^b:=*(\vt_a \wedge *\vt^b)=\d^b_a.
\label{2-6}
\end{equation}
Note that the product defined by (\ref{2-5}) is really a multiplication of two exterior 
forms thus this type of definition justifies the term ``interior product''.
We use in this paper the definition of the interior product in the form (\ref{2-5}).\\
The following first-order differential operators are defined on a teleparallel manifold:
\begin{itemize}
\item[\bf{1}] The exterior derivative operator $d:\Omega^p\to\Omega^{p+1}$
\begin{equation}
d\a=d\Big(\a_{a_1\cdots a_p}dx^{a_1}\wedge\cdots\wedge dx^{a_p}\Big)=
d\a_{a_1\cdots a_p}\wedge dx^{a_1}\wedge\cdots\wedge dx^{a_p}.
\label{2-7}
\end{equation}
This is an anti-derivative relative to the wedge product of forms
 \begin{equation}
d(\a\wedge\b)=d\a\wedge\b+(-1)^{deg\a}\a\wedge d\b.
\label{2-7a}
\end{equation}
\item[\bf{2}] The coderivative operator $d:\Omega^p\to\Omega^{p-1}$ defined by
\begin{equation}
d^+\a=*d*\a.
\label{2-8}
\end{equation}
\end{itemize}
Using the operators described above the action functional can be accepted on the 
teleparallel manifold in the quadratic (Dirichlet) form.  
The simplest choice of the Lagrangian density can be made in the form of the Yang-Mills 
Lagrangian:
\begin{equation}
S[\vt^a] = \int_M  d\vt^a \wedge *d\vt_a.
\label{2-9}
\end{equation}
Observe that the action (\ref{2-9}) satisfies the following conditions: 
\begin{itemize}
\item[\bf{1}] It is independent on a particular choice of a local coordinate system.
\item[\bf{2}] It includes only the first order derivatives 
of the field variables (coframe 1-forms) 
so the corresponding field equation is at most of the second order.
\item[\bf{3}] It is a functional of a Dirichlet-type so it provides an equation of
a harmonic type.
\item[\bf{4}] It is invariant under the global group of $SO(1,3)$ transformations 
of the coframe field $\vt^a$.
\end{itemize}
Therefore the generalization of the pseudo-Riemannian structure to a teleparallel structure 
allows to consider general Lagrangians for gravity similar to the quadratic Lagrangians 
of the other (already quantized) field theories. In this way one can hope to define the energy 
of gravity field in a local, covariant form. 
\sect{Variational procedure}
The variational procedure on a teleparallel manifold for the action  functional 
$S=S[\vt^a]$  can be described as follows. \\
Let $\vt^a(\lambda)$  be a smooth 1-parametric family of cross-sections 
of $FM$, with the following initial conditions 
\begin{equation}
\vt^a(\la=0)=
\vt^a\hbox{\qquad and\qquad}{\frac{\partial\vt^a}{\partial\la}}\bigg|_{\la=0}=\delta\vt^a.  
\label{3-1}\end{equation} 
The critical points of the functional $S[\vt^a]$ are 
 defined by the condition:
\begin{equation}
\delta S={\frac{dS}{d\lambda}} \bigg |_{\lambda=0} d\la=0.
\label{3-2}\end{equation}
The Hamilton principle of the least action postulates that the field equation of the 
physical system coincides with the critical point condition for an appropriative action 
functional. \\
The variation $\delta$ ($\lambda$-differential) is independent 
on the space-time coordinates so it satisfies the following rules.
\begin{itemize}
\item[\bf{1}]The ordinary Leibniz rule for the wedge product
\begin{equation}
\delta(\a \wedge \b)=\delta\a \wedge \b+\a \wedge \delta\b.
\label{3-3}\end{equation}
\item[\bf{2}] The commutativity with the exterior derivative
\begin{equation}
\delta \circ d=d \circ \delta
\label{3-4}\end{equation} 
\item[\bf{3}] The non-commutativity (in general) with the Hodge star-operator
\begin{equation}
\delta \circ *\ne * \circ \delta
\label{3-5}\end{equation}
\item [\bf{4}] The non-commutativity (in general) with the coderivative operator
\begin{equation}
\delta \circ d^+\ne d^+ \circ \delta
\label{3-6}\end{equation}
\end{itemize}
The following lemma is useful for the actual calculations of the variations.
\lem A variation of the wedge product monomials satisfies         
 \begin{equation}
\d\vt^{a_1...a_p} = \d\vt^m \wedge (\vt_m \hook \vt^{a_1...a_p}),
\label{3-7}\end{equation}
and for the Hodge dual monomials
\begin{equation}
\d*\vt^{a_1...a_p} = \d\vt^m \wedge (\vt_m \hook *\vt^{a_1...a_p}).
\label{3-8}\end{equation}
\elem
\prf
 Using the Leibniz rule (\ref{3-3}) we obtain
\begin{equation}
 \d\vt^{a_1...a_p} = \d\vt^{a_1} \wedge \vt^{a_2...a_p}+\vt^a_1 \wedge \d\vt^{a_2} \wedge \vt^{a_3...a_p}+...+\vt^{a_1...a_{p-1}} \wedge \d\vt^{a_p}.
\label{3-9}\end{equation}
The property of the interior product yields
\begin{equation}
\vt_m\hook\vt^{a_1\cdots a_p}=\d^{a_1}_m\vt^{a_2\cdots a_p}-\d^{a_2}_m\vt^{a_1a_3\cdots a_p}+\cdots.
\label{3-10}\end{equation}
Thus
\begin{equation}
\d\vt^m\wedge\Big(\vt_m\hook\vt^{a_1\cdots a_p}\Big)=
\d\vt^{a_1} \wedge \vt^{a_2...a_p}-\d\vt^{a_2}\wedge\vt^{a_1a_3\cdots a_p}+\cdots.
\label{3-11}\end{equation}
Comparing (\ref{3-9}) and (\ref{3-11}) we obtain the required formula (\ref{3-7}). 
The second formula (\ref{3-8}) is the consequence of the previous one, because the Hodge 
dual of a wedge product monomial is also a monomial.
\eprf \\
Therefore, we have the following formula for variation of the dual form
\begin{equation}
\d*\vt^a=\d\vt_m\wedge *(\vt^m\wedge\vt^a)
\label{3-12}\end{equation}
Thus the variation of the dual form is essentially differ from the 
variation $\d\vt^a$ of the basis form $\vt^a$.\\
For the variation of the volume element the formula (\ref{3-8}) yields
 \begin{equation}
\d*1=\d\vt^m \wedge (e_m \hook *1)=\d\vt^m \wedge *(\vt_m\wedge *^2 1)=
-\d\vt^m \wedge *\vt_m.
\label{3-13}\end{equation}
\pro
For an arbitrary differential form $\a \in \Omega^p$ on a $n$-dimensional 
manifold the following commutative relation holds
\begin{equation}
\d*\a - *\d\a=
*\Big(\d\vt_m\hook(\vt^m\wedge \a)-\d\vt_m\wedge(\vt^m\hook\a)\Big).
\label{3-14}\end{equation}
\epro
\prf
Evaluating the $p$-form $\a$ in the basis component forms
$$
\a=\a_{a_1...a_p}\vt^{a_1...a_p}
$$
and using the relation (\ref{3-8}) we derive 
\begin{eqnarray*}
\d*\a&=&
\d(\a_{a_1...a_p})*\vt^{a_1...a_p}+\a_{a_1...a_p}\d*\vt^{a_1...a_p}\\
&=&\d(\a_{a_1...a_p})*\vt^{a_1...a_p}+
\a_{a_1...a_p}\d\vt^m\wedge(e_m\hook *\vt^{a_1...a_p})\\
&=&\d(\a_{a_1...a_p})*\vt^{a_1...a_p}+\d\vt^m\wedge(\vt_m\hook *\a).
\end{eqnarray*}
On the other hand
\begin{eqnarray*}
\d\a&=&
\d(\a_{a_1...a_p})\vt^{a_1...a_p}+\a_{a_1...a_p}\d(\vt^{a_1...a_p})\\
&=&\d(\a_{a_1...a_p})\vt^{a_1...a_p}+\d\vt^m\wedge(\vt_m\hook *\a).
\end{eqnarray*}
Therefore 
\begin{eqnarray*}
\d*\a-*\d\a&=&
\d\vt_m\wedge(\vt^m\hook *\a)-*\Big(\d\vt_m\wedge(\vt^m\hook *\a)\Big)\\
&=&\Big((-1)^{p(n-p)+1}\d\vt_m\wedge*(\vt^m\wedge\a)-
*(\d\vt_m\wedge(\vt^m\hook\a))\Big)\\
&=&*\Big[\d\vt_m\hook(\vt^m\wedge\a)-\d\vt_m\wedge(\vt^m\hook\a)\Big]
\end{eqnarray*}
This proves the relation (\ref{3-14}).
\eprf\\
It is easy to see from the proof of the proposition above that in a particular case of
basis forms the following (anti)commutativity relations hold
\begin{equation}
\d*\vt^{a_1...a_p} \pm *\d\vt{a_1...a_p} =
*\Big(\d\vt_m\hook\vt^{ma_1...a_p}\pm
\d\vt_m\wedge(\vt^m\hook\vt^{a_1...a_p})\Big).
\label{3-15}
\end{equation}
The commutative relation similar to (\ref{3-14}) was proposed for the first
time in \cite{Hehl}.
The slightly different form is exhibited in \cite{BFK}.
The relation (\ref{3-14}) coincides
with the master formula of \cite{Hehl} in a  case of a pseudo-orthonormal
coframe.

\sect{Variations and constrains.}
The ordinary variational problem is used to study the critical points
of a real-valued functional. The variations of the field variables
are considered to be independent - free variations. In order to study
a constraint physical system, for instant a motion of a particle on a
fixed surface, one consider a critical point of a functional restricted
by appropriative constrains. This restriction is usually incorporated
by the method of Lagrangian multipliers. Another view on the constrained
variational problem can be proposed by consideration of constraint
variations. These variations satisfy some additional conditions.
In the framework of a global $SO(1,3)$ invariant and diffeomorphic
covariant theory we are interesting in covariant constraint conditions.
Denote the variation of the basis 1-form $\vt^a$ by
\begin{equation}
\d\vt^a=\d\vt^a={\epsilon^a}_b\vt^b.
\label{4-1}\end{equation}
We will refer to the matrix $\epsilon_{ab}=\eta_{ac}{\epsilon^c}_b$
as the  \textsl{ variational matrix.}\sf\\
The variational matrix admits a natural covariant decomposition
\begin{equation}
\epsilon_{ab}=\epsilon^{(1)}_{ab}+\epsilon^{(2)}_{ab}+\epsilon^{(3)}_{ab},
\label{4-2}\end{equation}
where 
\begin{eqnarray}
\epsilon^{(1)}_{ab}&=&{\frac{1}{2}}(\epsilon_{ab}-\epsilon_{ba})
\qquad \qquad \textsl{ - antisymmetric variation,}\sf\label{4-3}\\
\epsilon^{(2)}_{ab}&=&{\frac{1}{2}}(\epsilon_{ab}+\epsilon_{ba}) -
\epsilon \eta_{ab}
\qquad \textsl{- traceless symmetric variation,}\sf\label{4-4}\\
\epsilon^{(2)}_{ab}&=&\epsilon \eta_{ab} \qquad \qquad \qquad \qquad
\textsl{ {- trace variation.}}\sf\label{4-5}
\end{eqnarray}
We use here and later the notation $\epsilon={\epsilon^a}_a=\epsilon_{ab}\eta^{ab}$ 
for a trace of the variation matrix.
We can study now a free variation with an arbitrary
variational matrix as well as  various constrain variations,
determined by a special choice of the variational matrix $\epsilon_{ab}$.\\
Let us examine the conditions given by the different constrains.\\
The question is: What condition should be imposed on the variations of
$\{\vt^a\}$ in order to obtain some commutativity conditions with
the Hodge star operator?\\
Consider two following cases
\begin{equation}
\d*\vt^a=*\d\vt^a   \qquad \qquad\quad \textsl{{- commuatativity}}
\sf\label{4-6}\end{equation}
\begin{equation}
\d*\vt^a=-*\d\vt^a    \qquad \qquad \textsl{{- anticommuatativity}}
\sf \label{4-7}\end{equation}
Using (\ref{3-12}) in the right hand sides of (\ref{4-6},\ref{4-7})
and   considering  they together we obtain
\begin{equation}
*\d\vt^a=\pm\d\vt_m\wedge *(\vt^m\wedge\vt^a).
\label{4-8}\end{equation}
Hence the (anti)commutativity  conditions take the form
\begin{equation}
*\d\vt^a=\pm\d\vt_m\hook\vt^{ma}.
\label{4-9}\end{equation}
Inserting the variational matrix notation we derive
\begin{eqnarray*}
{\epsilon^a}_b\vt^b&=&\pm{\epsilon_m}^k\vt_k\hook\vt^{ma}
=\pm{\epsilon_m}^k(\d^m_k\vt^a-\d^a_k\vt^m)\\
&=&\pm({\epsilon_m}^m\vt^a-{\epsilon_m}^a\vt^m)=
\pm(\epsilon d^a_b-{\epsilon_b}^a)\vt^b.
\end{eqnarray*}
Therefore in the notations of the variational matrix the
(anti)commutativity  conditions (\ref{4-9}) take the form
\begin{equation}
{\epsilon^a}_b = \pm(\epsilon\d^a_b-{\epsilon_b}^a).
\label{4-10}\end{equation}
Taking the trace in two sides of this matrix equation we obtain
$$
\epsilon=\pm 3\epsilon.
$$
Thus the first necessary condition for (anti)commutativity of the variation
with the Hodge dual is
\begin{equation}
\epsilon=0.
\label{4-11}
\end{equation}
The equation (\ref{4-10}) gives the second condition for (anti)commutativity
\begin{equation}
\epsilon_{ab}\pm\epsilon_{ba}=0.
\label{4-12}\end{equation}
The consideration above can be summarized in the following
\pro\\
The variation commutes with the Hodge dual of a basis 1-form
\begin{equation}
\d*\vt^a=*\d\vt^a
\label{4-13}\end{equation}
if and only if the variational matrix $\epsilon^{ab}$ is antisymmetric.\\
The variation anticommutes with the Hodge dual of a basis 1-form
\begin{equation}
\d*\vt^a=-*\d\vt^a
\label{4-14}\end{equation}
if and only if the variational matrix
$\epsilon^{ab}$ is traceless and symmetric.
\epro
Consider now the variation of the metric tensor
$$
\d g_{\mu \nu}=\d(\eta_{ab}\vt^a_\mu\vt^b_\nu)=
\eta_{ab}(\d\vt^a_\mu\vt^b_\nu +\vt^a_\mu \d\vt^b_\nu).
$$
But
$$
\d\vt^a={\epsilon^a}_b\vt^b=\d(\vt^a_\mu dx^\mu)=
\d\vt^a_\mu dx^\mu=\vt^\mu_b\d\vt^a_\mu\vt^b.
$$
Therefore
$$\vt^\mu_b\d\vt^a_\mu={\epsilon^a}_b,$$
or
$$\d\vt^a_\mu={\epsilon^a}_b\vt^b_\mu.$$
Hence  the variation of the metric tensor is
\begin{equation}
\d g_{\mu \nu}=\eta_{ab}
({\epsilon^a}_c\vt^c_\mu\vt^b_\nu+{\epsilon^b}_c\vt^a_\mu\vt^c_\nu)=
\epsilon_{bc}(\vt^c_\mu\vt^b_\nu+\vt^b_\mu\vt^c_\nu).
\label{4-15}\end{equation}
Note that the expression in the brackets is symmetric under the permutation
of the indices $c$ and $b$.
Thus the variation of the metric tensor $\d g_{\mu \nu}$ is identically
equal to zero if and only if the matrix $\epsilon_{bc}$ is antisymmetric. 
Hence the commutativity of the variational operator with the Hodge dual map
results in a zero variation of the metric tensor.\\
In the second (anti-commutativity) case the variation of the metric tensor
is generally differ from zero.
This variation is traceless and symmetric so it represents a
9-parametric family of transformations. \\
The traceless variations preserve the determinant of the metric tensor
$g^{\mu\nu}$. Indeed the relation (\ref{4-15}) yields:
\begin{equation}
g^{\mu\nu}\d g_{\mu\nu}=2\epsilon_{bc}\eta^{bc}=2\epsilon.
\label{4-16}\end{equation}
The variation of the determinant of the metric tensor can be written as
\begin{equation}
\d g=gg^{\mu\nu}\d g_{\mu\nu}.
\label{4-17}\end{equation}
Therefore in the case of the traceless variations
\begin{equation}
\d g=2\epsilon g=0.
\label{4-18}\end{equation}
Note a connected result.
Since $\d*1=\epsilon *1$ it follows that the volume element is preserved
if and only if $\epsilon=0$. Summarize the conclusions in the following table:

\vspace{0.3cm}

\newpage
\qquad {\bf{ Constrained variations and commutativity rules.}}

\vspace{0.4cm}

\begin{tabular}{|c|c|c|c|c|}
\hline Type of&Variational&Metric &Variation&Commutativity \\
variation & matrix &tensor&of volume&rules \\ 
\hline & & & & \\
free & $\epsilon_{ab}$ &$\d g^{\mu\nu}\ne 0$ &$ \d*1 \ne 0$& \\
\hline & & & & \\
antisymmetric&$\epsilon^{(1)}_{ab}$&$\d g^{\mu\nu}=0$&$\d*1=0$&$\d*\vt^a=*\d\vt^a$ \\
\hline & & & & \\
symmetric&$\epsilon^{(2)}_{ab}+\epsilon^{(3)}_{ab}$&$\d g^{\mu\nu}\ne 0$&$\d*1\ne0$&\\
\hline & & & &\\
traceless&$\epsilon^{(1)}_{ab}+\epsilon^{(2)}_{ab}$&$\d g^{\mu\nu}\ne 0$&$\d*1=0$&\\
\hline traceless& & & &\\
symmetric&$\epsilon^{(2)}_{ab}$&$\d g^{\mu\nu}\ne 0$&$\d*1=0$&$\d*\vt^a=-*\d\vt^a$\\
\hline 
\end{tabular} 

\vspace{0.3cm}

We have established  the commutativity conditions only for the variations  of
the basis 1-forms, but they can be extended straightforward for basis
forms of an arbitrary degree. For instant, for a form $\a$ of an arbitrary
degree the formula (\ref{3-14}) yields
\begin{eqnarray*}
\d*\a-*\d\a&=&*{\epsilon_m}^k\Big(\vt_k\hook(\vt^m\wedge\a)-
\vt_k\wedge(\vt^m\hook\a)\Big)\\
&=&*{\epsilon_m}^k\Big(\d^m_k\a-\vt^m\wedge(\vt_k\hook\a)-
\vt_k\wedge(\vt^m\hook\a)\Big)\\
&=&*\epsilon\a-*\epsilon_{mk}\Big(\vt^k\hook(\vt^m\wedge\a)+
\vt^k\wedge(\vt^m\hook\a)\Big)
\end{eqnarray*}
The first term in the last expression vanish in the case of
a traceless variation. The expression in the brackets is
symmetric under the underchange of the indices $k$ and $m$.
Thus in the case of an antisymmetric variational matrix  $\epsilon_{mk}$
the variation commutes with the Hodge dual.
In the case of a basic form of an arbitrary degree
the relation (\ref{3-15}) yields to the anti-commutative condition
for traceless symmetric variation
\begin{equation}
\d*\vt^{a_1...a_p}=-*\d\vt^{a_1...a_p}.
\end{equation}
We see that the commutativity (anti-commutativity) of the variation
with the Hodge dual operator is connected with a certain special case of
constrained variation of the coframe field. The special case
of commutativity was proved for the first time in \cite{Hehl}.
\sect{General quadratic Lagrangian}
A typical Lagrangian used in the field theory is quadratic.
It is useful to have a formula for a variational derivative of a 
such type expression.
\pro
For $\a,\b \in \Omega^p$ on a $n$-dimensional manifold the variation 
of the square-norm expression $\L=\a\wedge*\b$ takes a following form
\begin{equation}
\d(\a\wedge*\b)=\d\a \wedge*\b +\a\wedge*\d\b-\d\vt_m \wedge J^m,
\label{5-1}\end{equation}
where 
\begin{equation}
J^m=(\vt^m\hook\b)\wedge *\a-(-1)^p\a\wedge (\vt^m\hook*\b).
\label{5-2}\end{equation}
\epro
\prf
Using the Leibniz rule for the variation (\ref{3-3}) we write
\begin{equation}
\d(\a\wedge*\b)=\d\a \wedge*\b +\a\wedge \d*\b.
\label{3-17}\end{equation}
The formula for commutator  (\ref{3-14}) yields
\begin{equation}
\d(\a\wedge*\b)=\d\a \wedge*\b +\a\wedge*\d\b +
\a\wedge *\Big(\d\vt_m\hook(\vt^m\wedge \b)-\d\vt_m\wedge(\vt^m\hook\b)\Big).
\label{3-18}\end{equation}
Compute the third term on the right hand side of (\ref{3-18}) 
\begin{eqnarray*}
\a\wedge *\Big(\d\vt_m\hook(\vt^m\wedge \b)\Big)
&=&\a\wedge *^2\Big(\d\vt_m\wedge *(\vt^m\wedge \b)\Big)\\
&=&(-1)^{p(n-p)+1}\a\wedge\d\vt_m\wedge *(\vt^m\wedge \b)\\
&=&(-1)^p\d\vt_m\wedge\a\wedge (\vt^m\hook*\b).
\end{eqnarray*}
The fourth term on the right hand side of (\ref{3-18}) is
\begin{eqnarray*}
\a\wedge *\Big(\d\vt_m\wedge(\vt^m\hook\b)\Big)
&=&\d\vt_m\wedge(\vt^m\hook\b)\wedge *\a.
\end{eqnarray*}
These prove the statement.
\eprf\\
The $n-1$ form $J^m$ represents a certain type of a field current. 
Because of a symmetry of the expression (\ref{5-1}) under the permutation of 
 $\a$ and $\b$ we can rewrite  the value $J^m$ in an explicitly 
symmetric form:
\begin{eqnarray}
J^m&=&\frac 12\Big([(\vt^m\hook\b)\wedge *\a+(\vt^m\hook\a)\wedge *\b]\nonumber\\
&& \qquad -(-1)^p[\a\wedge (\vt^m\hook*\b)+\b\wedge (\vt^m\hook*\a)]\Big).
\label{3-20}\end{eqnarray}
In the special case $\a=\b$ we obtain
\begin{equation}
J^m=(\vt^m\hook\a)\wedge *\a-(-1)^p\a\wedge (\vt^m\hook*\a).
\label{3-21}\end{equation}
The $n-1$ form of current $J^m$ can be written component-wisely as
\begin{equation}
J^m=J^{mn}*\vt_n.
\label{3-22}\end{equation}
Using
\begin{equation}
*(\vt^k\wedge J^m)=J^{mn}*(\vt^k\wedge*\vt_n)
\label{3-23}\end{equation}
we obtain the explicit expression for the matrix $J^{mn}$
\begin{equation}
J^{mn}=*(\vt^n\wedge J^m).
\label{3-24}\end{equation}
The two indexed object $J^{mn}$ can be considered as an analog of the 
energy-momentum tensor. 
\pro The matrix $J^{mn}$ is symmetric
\begin{equation}
J^{mn}=J^{nm}.
\label{3-25}\end{equation}
\epro
\prf
For the proof we use a simpler expression (\ref{5-2}) for $J^{mn}$,
 but take in account 
that this object  is actually symmetric for a permutation of the forms $\a$ and $\b$.
Using (\ref{5-2},\ref{3-24}) we derive
\begin{eqnarray*}
J^{mn}&=&*(\vt^n\wedge J^m)=*\Big[\vt^n\wedge\Big((\vt^m\hook\b)\wedge *\a-
(-1)^p\a\wedge (\vt^m\hook*\b)\Big)\Big]\\
&=&(-1)^{p-1}*\Big((\vt^m\hook\b)\wedge(\vt^n\wedge*\a)+
(\vt^n\wedge\a)\wedge (\vt^m\hook*\b)\Big).
\end{eqnarray*}
Write 
\begin{equation}
\vt^n\wedge*\a=(-1)^{\sigma_1}*^2\vt^n\wedge*\a=(-1)^{\sigma_1}*(\vt^n\hook\a),
\label{3-26}\end{equation}
\begin{equation}
\vt^m\hook*\b=*(\vt^m\wedge *^2\b)=(-1)^{\sigma_2}*(\vt^m\wedge \b),
\label{3-27}\end{equation}
where $\sigma_1,\sigma_2$ are two integers.
Thus we obtain 
\begin{eqnarray}
J^{mn}&=&(-1)^{p-1}*\Big((-1)^{\sigma_1}(\vt^m\hook\b)\wedge*(\vt^n\hook\a)+\nonumber\\
 &&(-1)^{\sigma_2}(\vt^n\wedge\a)\wedge *(\vt^m\wedge \b)\Big).
\label{3-28}\end{eqnarray}
Now the symmetry of $J^{mn}$ in the form (\ref{3-28}) under the permutation  of the 
forms $\a$ and $\b$ 
yields the symmetry under the permutation of the indices $m$ and $n$. 
\eprf\\
Calculate the trace of the matrix  $J^{mn}$.
\begin{eqnarray}
{J^m}_m&=&*(\vt_m\wedge J^m)=*\Big[\vt_m\wedge \Big((\vt^m\hook\b)\wedge *\a-
(-1)^p\a\wedge(\vt^m\hook *\b)\Big)\Big]\nonumber\\
&=&*
\Big(\vt_m\wedge(\vt^m\hook\b)\wedge *\a-\a\wedge\vt_m\wedge(\vt^m\hook *\b)\Big).
\end{eqnarray}
Use the formula 
\begin{equation}
\vt_m\wedge(\vt^m\hook w)=p w,
\label{3-29}\end{equation}
where $w$ is an arbitrary $p$-form.
The trace takes the form
\begin{equation}
{J^m}_m=p\b\wedge *\a-(n-p)\a\wedge *\b=-(n-2p)*(\a\wedge *\b).
\label{3-30}\end{equation}
Thus we obtain
\pro
For a quadratic Lagrangian 
\begin{equation}
\L=\a\wedge*\b \qquad deg\a=deg\b=p
\label{3-31}\end{equation}
 on a manifold $M$ the matrix $J^{mn}$ is traceless ${J^m}_m=0$ if and only if 
the dimensional of the manifold $M$ is even.
\epro
The second condition, namely $deg\a=deg\b=\frac n2$, does not so crucial as 
it can be supposed on a first look. Indeed,  for  a 4-dimensional 
manifold three different cases are possible
\begin{eqnarray}
\label{3-31a}
deg\a&=&deg\b=2,\\
\label{3-31b}
deg\a&=&deg\b=1,\\
\label{3-31c}
deg\a&=&deg\b=2.
\end{eqnarray}
In the second case we can rewrite the Lagrangian as 
\begin{equation}
\L=\a\wedge*\b=\vt^a\wedge\a\wedge(\vt_a\hook*\b)=
(\vt^a\wedge\a)\wedge*(\vt_a\wedge\b).
\end{equation}
Now this is a product of two forms of a second degree. In the third case 
(\ref{3-31c}) we can apply the same transformation twice.\\
The variational procedure described above can be straightforward extended 
for a general situation with an external field on the manifold $\{M,\vt^a\}$.
Consider, for instance, the Lagrangian 
\begin{equation}
\L=\a(\phi,d\phi,\vt^a)\wedge*\b(\phi,d\phi,\vt^a),
\label{3-21a}\end{equation}
where $\phi$ is a certain external field (possible multicomponent) - 
differential form of degree $q$. 
Write the variation of the forms $\a,\b$ as
 \begin{equation}
\d\a=\d\phi\wedge\frac{\d\a}{\d\phi}+\d d\phi\wedge\frac{\d\a}{\d d\phi}+
\d\vt^m\wedge\frac{\d\a}{\d\vt^m},
\label{3-22a}\end{equation}
 \begin{equation}
\d\b=\d\phi\wedge\frac{\d\b}{\d\phi}+\d d\phi\wedge\frac{\d\b}{\d d\phi}+
\d\vt^m\wedge\frac{\d\b}{\d\vt^m}.
\label{3-23a}\end{equation}
The relations  (\ref{5-1}) yields
\begin{eqnarray}
\d\L&=&(\d\phi\wedge\frac{\d\a}{\d\phi}+\d d\phi\wedge\frac{\d\a}{\d d\phi}+
\d\vt^m\wedge\frac{\d\a}{\d\vt^m}) \wedge*\b +\nonumber\\
&&
(\d\phi\wedge\frac{\d\b}{\d\phi}+\d d\phi\wedge\frac{\d\b}{\d d\phi}+
\d\vt^m\wedge\frac{\d\b}{\d\vt^m})\wedge*\a
-\d\vt_m \wedge J^m.
\label{3-24a}\end{eqnarray}
Thus we obtain two Euler-Lagrange field equations
\begin{equation}
\frac{\d\a}{\d\phi}\wedge*\b+\frac{\d\b}{\d\phi}\wedge*\a-
(-1)^q d\Big(\frac{\d\a}{\d d\phi}\wedge*\b+
\frac{\d\b}{\d d\phi}\wedge*\a\Big)=0,
\label{3-26a}\end{equation}
\begin{equation}
\frac{\d\a}{\d\vt^m} \wedge*\b +\frac{\d\b}{\d\vt^m}\wedge*\a=J^m.
\label{3-27a}\end{equation}
In two following sections we apply the variational procedure described 
above for two important cases: the Maxwell Lagrangian for electromagnetism and 
the Rumpf Lagrangian for translation invariant gravity.
\sect{Maxwell Lagrangian on a teleparallel space}
The Lagrangian of the Maxwell theory of electromagnetism can be written as
\begin{equation}
\L=dA\wedge*dA,
\label{6-1}\end{equation}
where $A$ is a 1-form of electro-magnetic potential. This form of 
Lagrangian represents the classical theory of electromagnetism on a flat Minkowskian space.\\
If one consider the field $A$ to be defined on a curved pseudo-Riemannian 
manifold the 4-form $\L$ is well defined (invariant under the transformation 
of coordinates). The Hodge star operator in this situation depends on 
the metric $g$ on the manifold. 
Thus it is better to write the electro-magnetic Lagrangian on
a pseudo-Riemannian manifold as 
\begin{equation}
\L=dA\wedge*_gdA.
\label{6-2}\end{equation}
The variation of the Lagrangian have to be applied 
by using a free variation of the electro-magnetic 
potential $A$  as well as a free variation of the metric $g$.\\
On a teleparallel manifold the Lagrangian can be taken in the same form, 
but now the Hodge star operator depends on the coframe field $\vt^a$
\begin{equation}
\L=dA\wedge*_{\vt^a}dA.
\label{6-3}\end{equation}
The variation of the Lagrangian (\ref{6-3}) can be produced using the formulas
(\ref{5-1}, \ref{5-2}).
\begin{equation}
\d\L=\d(dA)\wedge*dA+dA\wedge*\d(dA)-\d\vt_m \wedge J^m,
\label{6-4}\end{equation}
where
\begin{equation}
J^m=(\vt^m\hook dA)\wedge *dA-dA\wedge (\vt^m\hook*dA).
\label{6-5}\end{equation}
Extracting the total derivative in (\ref{6-4}) 
\begin{eqnarray*}
\d\L&=&2d(\d A)\wedge*dA-\d\vt_m \wedge J^m\\&=&
2d(\d A\wedge*dA)+2\d A\wedge d*dA-\d\vt_m \wedge J^m
\end{eqnarray*}
we obtain two field equations
\begin{equation}
d*dA=0
\label{6-6}\end{equation}
and
\begin{equation}
J^m=0,
\label{6-6a}\end{equation}
where
\begin{equation}
J^m=(\vt^m\hook dA)\wedge *dA-dA\wedge (\vt^m\hook*dA).
\label{6-7}\end{equation}
The equation (\ref{6-6}) has exactly the same  form as the ordinary electro-magnetic field 
equation, but it is an equation on a curved manifold (operator $*$ depends 
on the coframe $\vt^a$). For interpretation the equations  
(\ref{6-6},\ref{6-6a}) introduce the strength notation for electro-magnetic
field
\begin{equation}
dA=F.
\label{6-8}\end{equation}
The first field equation (\ref{6-6}) takes the form of ordinary 
Maxwell equations
\begin{eqnarray}
dF&=&0,\\
\label{6-9}
d*F&=&0.
\label{6-10}
\end{eqnarray}
The equation (\ref{6-6a}) gives an additional relation
\begin{equation}
J^m=(\vt^m\hook F)\wedge *F-F\wedge (\vt^m\hook*F).
\label{6-10a}\end{equation}
The 2-form $F$ can be explained component-wisely as
\begin{equation}
F=F_{mn}\vt^{mn}.
\label{6-8a}\end{equation}
Note that the coefficients of $F$ are antisymmetric by definition 
$F_{mn}=-F_{nm}$. 
Consider the first term of $J^m$ (\ref{6-10a}) 
\begin{eqnarray*}
(\vt^m\hook F)\wedge *F&=&
(\vt^m\hook F^{kn}\vt_{kn})\wedge *F_{pq}\vt^{pq}\\ 
 &=&2F^{kn}F_{pq}\d^m_k\vt_n\wedge *\vt^{pq}=4F^{mn}F_{nq}*\vt^q
\end{eqnarray*}
The second term of (\ref{6-10a}) takes the form
\begin{eqnarray*}
F\wedge (\vt^m\hook*F)&=&F^{kn}\vt_{kn}\wedge(\vt^m\hook*F_{pq}\vt^{pq})
=-F^{kn}F_{pq}\vt_{kn}\wedge*\vt^{mpq}\\&=&
F^{kn}F_{pq}\vt_{k}\wedge*(\d^n_m\vt^{pq}-2\d^n_p\vt^{mq})\\&=&
F^{kn}F_{pq}*(2\d^n_m\d^k_p\vt^{q}-2\d^n_p\d^k_m\vt^{q}+
2\d^n_p\d^k_q\vt^{m})\\&=&4F^{pm}F_{pq}*\vt^{q}+2F^{pq}F_{pq}*\vt^{m}
\end{eqnarray*}
Thus the object $J^m$ takes the form
\begin{equation}
J^m=8(F^{mn}F_{nk}-\frac 14 F^{pq}F_{pq}\d^m_k) *\vt^k.
\label{6-11}\end{equation}
or in a component-wise form
\begin{equation}
{J^m}_q=8(F^{mn}F_{nk}-\frac 14 F^{pq}F_{pq}\d^m_k).
\label{6-12}\end{equation}
We see now what is the field  equation  (\ref{6-7}) mean. 
It is a  vanishing condition for the trace of the matrix ${J^m}_q$. 
Note that due to the propositions (5.3-5.4) the traceless condition 
as well as the symmetric condition are the consequences of the definition 
of $J_{ab}$. Thus in the case of the Maxwell Lagrangian the object ${J^m}_n$ 
coincides (up to a numerical coefficient equal to $-\frac{1}{32\pi}$) with the classical 
expression of the energy-momentum tensor. Note that by the variational 
procedure described above we obtain it in a symmetric and a traceless form. 
\sect{Translational Lagrangian in gravity}
In the teleparallel approach to gravity the coframe field $\vt^a$ is the basic 
gravitational field variable. A general Lagrangian density for the 
coframe field $\vt^a$ (quadratic in the first order derivatives)
described by 
the gauge invariant translation Lagrangian of Rumpf \cite{Rumpf}.
Up to the $\Lambda$-term it can be written as 
\begin{equation}
\L=\frac 1{2\ell^2}\sum^3_{I=1}\rho_I{}^{(I)}V,
\label{7.1}
\end{equation}
where
\br
\label{7.1a}
{}^{(1)}\L&=&d\vt^a\wedge*d\vt_a,\\
\label{7.2}
{}^{(2)}\L&=&(d\vt_a \wedge \vt^a ) \wedge*( d\vt_b \wedge \vt^b),\\
\label{7.3}
{}^{(3)}\L&=&(d\vt_a \wedge\vt^b ) \wedge * \Big(d\vt_b \wedge \vt^a \Big).
\er
and $\ell$ is the Plank length constant. \\
Note that the Lagrangian (\ref{7.1}) is a linear combination of three independent terms every one of which is  
of the prescribed form $\a\wedge *\b$ and we can apply the procedure described 
above. 
Using the proposition (5.1) we obtain for the first Lagrangian (\ref{7.2})
\begin{eqnarray}
\d\Big({}^{(1)}\L\Big)&=&2\d(d\vt_a)\wedge*d\vt^a-\d(\vt_m)\wedge {}^{(1)}J^m
\nonumber\\
&=&2d\Big(\d(\vt_a)\wedge*d\vt_a\Big)+
2\d(\vt_a)\wedge d*d\vt^a-\d(\vt_m)\wedge {}^{(1)}J^m,
\label{7.4}
\end{eqnarray}
where 
\begin{eqnarray}
{}^{(1)}J^m&=&
(\vt^m\hook d\vt_a)\wedge *d\vt^a - d\vt^a\wedge (\vt^m\hook *d\vt_a)\nonumber\\
&=&2(\vt^m\hook d\vt_a)\wedge *d\vt^a-\vt^m\hook(d\vt_a\wedge*d\vt^a)
\label{7.5}
\end{eqnarray}
Thus the contribution of the Lagrangian ${}^{(1)}\L$ in the field equation is
\begin{equation}
\boxed{2\rho_1\d(\vt_a)\wedge d*d\vt^a+2\rho_1(\vt^a\hook d\vt_m)\wedge *d\vt^m-
\rho_1\vt^a\hook(d\vt_m\wedge*d\vt^m)}
\label{7.6}
\end{equation}
Variation of the second Lagrangian (\ref{7.3}) takes the form
\begin{eqnarray}
\d\Big({}^{(2)}\L\Big)&=&2\d(d\vt_a \wedge \vt^a)\wedge*(d\vt_b \wedge \vt^b)-
\d(\vt_m)\wedge {}^{(2)}J^m\nonumber\\
&=&2d(\d\vt_a) \wedge \vt^a\wedge*(d\vt_b \wedge \vt^b)
+2d\vt_a \wedge \d\vt^a\wedge*(d\vt_b \wedge \vt^b)-\nonumber\\
&&\d(\vt_m)\wedge {}^{(2)}J^m\nonumber\\
&=&2d\Big(\d\vt_a \wedge \vt^a\wedge*(d\vt_b \wedge \vt^b)\Big)
+2\d\vt_a \wedge d\Big(\vt^a\wedge*(d\vt_b \wedge \vt^b)\Big)\nonumber\\
&&
+2\d\vt^a\wedge d\vt_a \wedge*(d\vt_b \wedge \vt^b)-
\d(\vt_m)\wedge {}^{(2)}J^m\nonumber\\
&=&2d\Big(\d\vt_a \wedge \vt^a\wedge*(d\vt_b \wedge \vt^b)\Big)-
2\d\vt_a \wedge \vt^a\wedge d*(d\vt_b \wedge \vt^b)+\nonumber\\
&&
4\d\vt^a\wedge d\vt_a \wedge*(d\vt_b \wedge \vt^b)-
\d(\vt_m)\wedge {}^{(2)}J^m,
\label{7.7}
\end{eqnarray}
where the current term
\begin{eqnarray}
{}^{(2)}J^m \eqq \Big(\vt^m\hook (d\vt_b \wedge \vt^b)\Big)\wedge
*(d\vt_a \wedge \vt^a)+
(d\vt_a \wedge \vt^a)\wedge \Big(\vt^m\hook*(d\vt_b \wedge \vt^b)\Big)
\nonumber\\
\eqq -\vt^m\hook\Big((d\vt^a\wedge\vt_a)\wedge*(d\vt^b\wedge\vt_b)\Big)+
2\vt^m\hook(d\vt^a\wedge\vt_a)\wedge*(d\vt^b\wedge\vt_b)
\nonumber\\
\eqq -\vt^m\hook\Big((d\vt^a\wedge\vt_a)\wedge*(d\vt^b\wedge\vt_b)\Big)+
2(\vt^m\hook d\vt^a)\wedge\vt_a\wedge*(d\vt^b\wedge\vt_b)\nonumber\\
\eee +2d\vt^m\wedge*(d\vt^b\wedge\vt_b)
\label{7.8}
\end{eqnarray}
Thus the contribution of the Lagrangian ${}^{(2)}\L$ in the field equation is
\begin{equation}
\boxed{
\begin{array}{ll}
&-2\rho_2\vt^a\wedge d*(d\vt_b \wedge \vt^b)
+\rho_2\vt^a\hook\Big((d\vt^m\wedge\vt_m)\wedge*(d\vt^n\wedge\vt_n)\Big)+\\
&2\rho_2 d\vt_a \wedge*(d\vt_b \wedge \vt^b)+
-2\rho_2(\vt^a\hook d\vt^m)\wedge\vt_m\wedge*(d\vt^n\wedge\vt_n)
\end{array}}
\label{7.9}
\end{equation}
As for the third laplacian (\ref{7.4}) by the proposition (5.1) we obtain
\begin{eqnarray}
\d\Big({}^{(3)}\L\Big)\eqq2\d(d\vt_a \wedge \vt^b)\wedge*(d\vt_b \wedge \vt^a)-
\d\vt_m\wedge {}^{(3)}J^m=-
\d\vt_m\wedge {}^{(3)}J^m+\nonumber\\
\eee 2d(\d\vt_a) \wedge \vt^b\wedge*(d\vt_b \wedge \vt^a)+
2\d\vt^b\wedge d\vt_a \wedge*(d\vt_b \wedge \vt^a)\nonumber\\
\eqq 2d\Big(\d\vt_a \wedge \vt^b\wedge*(d\vt_b \wedge \vt^a)\Big)-
2\d\vt_a \wedge \vt^b\wedge d*(d\vt_b \wedge \vt^a)+\nonumber\\
\eee
4\d\vt^b\wedge d\vt_a \wedge*(d\vt_b \wedge \vt^a)-
\d\vt_m\wedge {}^{(3)}J^m.
\label{7.10}
\end{eqnarray}
As for the current term
\begin{eqnarray}
{}^{(3)}J^m \eqq \Big(\vt^m\hook(d\vt_a \wedge \vt^b)\Big)\wedge *(d\vt_b \wedge \vt^a)+
(d\vt_a \wedge \vt^b)\wedge \Big(\vt^m\hook*(d\vt_b \wedge \vt^a)\Big)\nonumber\\
\eqq -\vt^m\hook\Big((d\vt_a \wedge \vt^b)\Big)\wedge *(d\vt_b \wedge \vt^a)
+2\vt^m\hook(d\vt_a \wedge \vt^b)\wedge *(d\vt_b \wedge \vt^a)\nonumber\\
\eqq -\vt^m\hook\Big((d\vt_a \wedge \vt^b)\Big)\wedge *(d\vt_b \wedge \vt^a)
+2(\vt^m\hook d\vt_a) \wedge \vt^b\wedge *(d\vt_b \wedge \vt^a)+\nonumber\\
\eqq 
2d\vt_a\wedge *(d\vt^m \wedge \vt^a)
\label{7-11}
\end{eqnarray}
Hence the contribution of the Lagrangian ${}^{(3)}\L$ in the field equation is
\begin{equation}
\boxed{
\begin{array}{ll}
&-2\rho_3 \vt^b\wedge d*(d\vt_b \wedge \vt^a)
+\rho_3\vt^a\hook\Big((d\vt_m \wedge \vt^n)\Big)\wedge *(d\vt_n \wedge \vt^m)\\
&-2\rho_3(\vt^a\hook d\vt_m) \wedge \vt^n\wedge *(d\vt_n \wedge \vt^m)
+2\rho_3d\vt_b\wedge *(d\vt^a \wedge \vt^b)
\end{array}}
\label{7-12}
\end{equation}
The relations (\ref{7.6},\ref{7.9},\ref{7-12}) yield the field equation 
generated 
by free variations of the Rumpf Lagrangian cf. Kopczy\'nski
\cite{Kop} in the form \cite{Hehl}.
\br\label{fe}
  - 2\ell^2\Sigma_a &=& 2\rho_1 d*d\vt_a
 - 2\rho_2 \vt_a \wedge d *(d\vt^b\wedge \vt_b)-
  2\rho_3\vt_b \wedge d*( \vt_a \wedge d \vt^b ) \nonumber \\
 && + \rho_1 \Big[ e_a \hook( d\vt^b \wedge * d\vt_b) 
 -2 ( e_a \hook d\vt^b ) \wedge *d\vt_b\Big] \nonumber \\
 && + \rho_2\Big[2 d\vt_a \wedge * ( d\vt^b \wedge \vt_b) 
 + e_a \hook ( d\vt^c \wedge\vt_c \wedge * ( d\vt^b \wedge\vt_b) )\nonumber\\
&&\qquad -2( e_a \hook d\vt^b) \wedge \vt_b \wedge *
  ( d\vt^c \wedge \vt_c)\Big] \nonumber \\
 && +\rho_3\Big[2 d\vt_b \wedge * ( \vt_a\wedge d \vt^b)
+ e_a \hook( \vt_c \wedge d\vt^b \wedge *( d\vt^c \wedge \vt_b )) \nonumber \\
&&\qquad
- 2( e_a \hook d\vt^b) \wedge \vt_c \wedge * ( d\vt^c \wedge\vt_b )\Big] ,
\er
where $\Sigma_a$ depends on matter fields.

\sect{Conclusions}
We discuss the variational procedure on a teleparallel manifold. The situation
is rather differ from the variation on a pseudo-Riemannian manifold.
We have reproved  a part of the propositions exhibited in the paper
\cite{Hehl} and have generalized some of them. The 
commutativity and anti-commutativity of the variation with the Hodge dual map
are related with the special covariant types of the constraint variation.
We derive a general relation for variation of the quadratic type Lagrangians
and apply it to the viable cases of the electro-magnetic Lagrangian and to
the translation invariant of Rumpf. The established formulas can be applied
for study the various material fields on teleparallel background.
We hope that these
techniques will also be useful  for Lagrangian formulation of the
dynamic of $p$-branes.

\vspace{1.3cm}
\bf{Acknowledgments}:\\ \sf
I am grateful to Prof. Shmuel Kaniel for constant support and interesting
discussions. My gratitude to Prof. Friedrich W.~Hehl and to his collaborates 
for useful comments and helpful remarks. This work is considerably influenced 
by their paper \cite{Hehl}.

\end{document}